\documentclass[a4paper,11pt]{article}

\usepackage{graphicx}
\usepackage{grffile}
\usepackage{pos}
\usepackage{braket}
\usepackage{comment}
\usepackage{bbm}

\hypersetup{pdftitle={PoS(Corfu2023): Hirasawa, Μ, SUSY effects on emergent spacetime
    in the type IIB matrix model}}
\usepackage[colorinlistoftodos,shadow]{todonotes}

\title{The effects of SUSY on the emergent spacetime in the Lorentzian type IIB matrix model}
\ShortTitle{The effects of SUSY in the Lorentzian type IIB matrix model}

\author*[a, b]{Mitsuaki Hirasawa}
\author[c]{Konstantinos N. Anagnostopoulos}
\author[d]{Takehiro Azuma}
\author[e]{Kohta Hatakeyama}
\author[f, g]{Jun Nishimura}
\author[c]{Stratos Papadoudis}
\author[h]{Asato Tsuchiya}

\affiliation[a]{Department of Physics ``Giuseppe Occhialini'', University of Milano-Bicocca,\\ Piazza della Scienza, 3, I-20126 Milano, Italy}

\affiliation[b]{Sezione di Milano Bicocca, Istituto Nazionale di Fisica Nucleare,\\ Piazza della Scienza, 3, I-20126 Milano, Italy}

\affiliation[c]{Physics Department, School of Applied Mathematical and Physical Sciences,\\ National Technical University, Zografou Campus, GR-15780 Athens, Greece}
\affiliation[d]{Institute for Fundamental Sciences, Setsunan University,\\ 17-8 Ikeda Nakamachi, Neyagawa, Osaka, 572-8508, Japan}

\affiliation[e]{Department of Physics, Kyoto University,\\ Kyoto 606-8502, Japan}

\affiliation[f]{KEK Theory Center, High Energy Accelerator Research Organization,\\ 1-1 Oho, Tsukuba, Ibaraki 305-0801, Japan}
\affiliation[g]{Graduate Institute for Advanced Studies, SOKENDAI,\\ 1-1 Oho, Tsukuba, Ibaraki 305-0801, Japan}
\affiliation[h]{Department of Physics, Shizuoka University,\\ 836 Ohya, Suruga-ku, Shizuoka 422-8529, Japan}

\emailAdd{mitsuaki.hirasawa@mib.infn.it}
\emailAdd{konstant@mail.ntua.gr}
\emailAdd{azuma@mpg.setsunan.ac.jp}
\emailAdd{kohta.hatakeyama@gauge.scphys.kyoto-u.ac.jp}
\emailAdd{jnishi@post.kek.jp}
\emailAdd{sp10018@central.ntua.gr}
\emailAdd{tsuchiya.asato@shizuoka.ac.jp}

\abstract{The Lorentzian type IIB matrix model is a promising candidate for a nonperturbative formulation of superstring
  theory. Recently we performed complex Langevin simulations
  by adding a Lorentz invariant mass term as an IR regulator
  and found a (1+1)--dimensional
  expanding spacetime with a Lorentzian signature emerging dynamically at late times
  when the fermionic contribution is omitted.
  Here
  we find that this is merely
  an artifact of the Lorentz boosts by showing that
  the spontaneous breaking of rotational symmetry is eliminated
  if one chooses a Lorentz frame appropriately.
  On the other hand, when we include the fermionic contribution,
  we find some evidence suggesting the emergence of a smooth
  (3+1)--dimensional expanding Lorentzian spacetime.}

\FullConference{Corfu Summer Institute 2023 "School and Workshops on Elementary Particle Physics and Gravity"\\
 23 April - 6 May , and 27 August - 1 October, 2023\\
Corfu, Greece\\}

\begin{document}

\begin{flushright}
KEK-TH-2636, KUNS-3007
\end{flushright}

\maketitle

\section{Introduction}
Superstring theory is a promising candidate for a quantum theory of gravity,
but it is consistently defined only in 10--dimensional
spacetime. The standard model of particle physics, on the other hand, is
defined in (3+1)--dimensional spacetime.
By compactifying the extra dimensions, the effective spacetime in superstring theory
at low energy becomes (3+1)--dimensional.
However, there are infinitely many perturbative vacua including those having
spacetime with different dimensions,
and we cannot determine the vacuum that corresponds to our universe,
at least at the perturbative level. Nonperturbative effects are thought to play
an important role in lifting this degeneracy.

The type IIB matrix model \cite{Ishibashi:1996xs}
is a promising candidate for a nonperturbative formulation of
superstring theory. The model is defined by dimensionally reducing 10--dimensional ${\cal N} = 1$ super Yang–Mills theory to zero
dimensions. Spacetime emerges dynamically from the bosonic matrix degrees of
freedom. The Euclidean version of this model has been studied analytically using the Gaussian expansion
method \cite{Nishimura:2001sx,Kawai:2002jk,Aoyama:2006rk,Nishimura:2011xy}
and also numerically using the complex Langevin
method \cite{Anagnostopoulos:2017gos,Anagnostopoulos:2020xai},
where the emergence of Euclidean 3--dimensional space has been
observed. However, the relationship between the emergent space and our (3+1)--dimensional universe was unclear. On the other hand,
the Lorentzian model has not been studied well due to the severe sign problem, which prevents us from applying conventional
Monte Carlo methods\footnote{ In Refs.~\cite{Kim:2011cr, Ito:2013ywa, Ito:2015mxa}, the Lorentzian model was studied by Monte
Carlo methods using an approximation and
the emergence of (3+1)--dimensional expanding spacetime was reported.
However, it was found later \cite{Aoki:2019tby}
that the approximation amounts to replacing
the complex weight $e^{iS}$ by $e^{-\beta S}$ ($\beta>0$),
and that the emergent space is actually not continuous.}.
Recently, numerical simulations have been performed using the complex Langevin method
\cite{Parisi:1983mgm,Klauder:1983sp} to overcome the sign
problem \cite{Nishimura:2019qal,Hirasawa:2021xeh,Hatakeyama:2021ake,Hatakeyama:2022ybs,Nishimura:2022alt,Hirasawa:2022qzg, Hirasawa:2023lpb}.
The hope is that the dynamics of the model will result in
the emergence of a
(3+1)--dimensional expanding spacetime, where the extra dimensions 
are compactified via a spontaneous symmetry breaking (SSB) of the 9--dimensional
rotational symmetry of space.
See Refs.~\cite{Anagnostopoulos:2022dak,Brahma:2022ikl,Klinkhamer:2022frp,Steinacker:2024unq}
for recent reviews on this model.

In our previous studies \cite{Nishimura:2022alt, Hirasawa:2022qzg, Hirasawa:2023lpb},
we demonstrated that the
Euclidean and the Lorentzian models are connected via analytic continuation. By adding a Lorentz invariant mass term in the
action \cite{Steinacker:2017vqw, Steinacker:2017bhb, Sperling:2018xrm, Sperling:2019xar, Steinacker:2019dii, Steinacker:2019awe,Steinacker:2019fcb, Hatakeyama:2019jyw, Steinacker:2020xph, Fredenhagen:2021bnw, Steinacker:2021yxt, Asano:2021phy,Karczmarek:2022ejn, Battista:2022hqn},
which acts as an IR regulator, the Lorentzian model becomes inequivalent to the Euclidean
model. In particular,
we provided evidence for obtaining a smooth expanding spacetime.
The signature of the metric changes
dynamically, being Euclidean at early times and becoming Lorentzian at late times.
While we found no evidence for a (3+1)--dimensional expanding spacetime,
we reported that by omitting the fermionic contribution and tuning
the model's parameters, a (1+1)--dimensional expanding spacetime emerges.  

In this paper, we first show that this is merely an artifact resulting from the action
of the Lorentz boost during the simulation due to the Lorentz symmetry of the model.
By choosing a Lorentz frame that offers a natural definition of spacetime,
the boost's effect is eliminated,
and the 1--dimensional expansion disappears.
Subsequently, we simulate the model by
incorporating the dynamical effect of the fermions and present evidence
suggesting the emergence of a smooth (3+1)--dimensional
expanding spacetime, with six dimensions of space compactified via
the SSB
of the
SO(9) rotational symmetry, in which
supersymmetry (SUSY) plays a crucial role.

The rest of this paper is organized as follows. In Section \ref{regularization}, we explain the regularization of the Lorentzian
model used in this work. In Section \ref{boost}, we present the results obtained by performing a simulation of the bosonic model
and point out that the appearance of the 1--dimensional expanding space is an artifact of the Lorentz boost. We also explain the
method for removing this artifact.  In Section \ref{SUSY}, we show our results obtained by simulations of the model including the
fermionic contribution.  Section \ref{summary} is devoted to a summary and discussions.

\section{Regularization of the Lorentzian type IIB matrix model}
\label{regularization}
The Lorentzian type IIB matrix model is defined by the partition function
\begin{equation}
    \begin{split}
      Z&=\int dA d\Psi
      \ e^{i(S_{\rm b} + S_{\rm f})} \ ,\\
        S_{\rm b}&=-\frac{N}{4} {\rm Tr}\left\{ -2[A_0, A_i]^2 + [A_i,A_j]^2 \right\} \ ,\\
        S_{\rm f}&=-\frac{N}{2} {\rm Tr}\left\{
        \Psi_\alpha
        (C\Gamma^\mu)_{\alpha\beta}[A_\mu,\Psi_{\beta}] \right\} \ ,
    \end{split}
    \label{partition-fn-original}
\end{equation}
where $A_\mu$ and $\Psi_\alpha$ are bosonic and fermionic $N\times N$ Hermitian
matrices, respectively. The indices $\mu$ and
$\alpha$ run from 0 to 9 and 1 to 16, respectively, while the spatial index $i$ runs from $1$ to $9$ only. The $16\times 16$
matrices $C$ and $\Gamma^\mu$ are the charge conjugation matrix and 10d Gamma matrices, respectively, after the Weyl projection.
This model has $\mathcal{N}=2$ SUSY, which is the maximal SUSY in 10d, implying that the model includes
gravity. Furthermore, from the SUSY algebra, one can identify the constant shift $A_\mu \to A_\mu + c_\mu \mathbbm{1}$ as the
translation in this model. Thus, the eigenvalues of the matrices $A_\mu$ can be identified as the spacetime coordinates. This model
also has SO(9,1) Lorentz symmetry,
which should be partially broken at some point in time
for the emergence of (3+1)--dimensional spacetime.

Since the partition function \eqref{partition-fn-original}
of the Lorentzian
model is not absolutely convergent,
we need to regularize it.
In this work, we use the Lorentz invariant mass term
\begin{equation}
    S_\gamma = -\frac{1}{2}N\gamma {\rm Tr}\left( A_\mu \right)^2 = \frac{1}{2}N\gamma \left\{ {\rm Tr}\left( A_0 \right)^2 - {\rm Tr}\left( A_i \right)^2 \right\}
\end{equation}
as an IR regulator,
where $\gamma$ is a mass parameter
and the $\gamma\to 0$ limit should be taken after taking the large--$N$ limit.
The model with this regulator has been studied in various
contexts \cite{Steinacker:2017vqw, Steinacker:2017bhb, Sperling:2018xrm, Sperling:2019xar, Steinacker:2019dii, Steinacker:2019awe, Steinacker:2019fcb, Hatakeyama:2019jyw, Steinacker:2020xph, Fredenhagen:2021bnw, Steinacker:2021yxt, Asano:2021phy, Karczmarek:2022ejn, Battista:2022hqn}. 
In particular, Ref.~\cite{Hatakeyama:2019jyw}
reported the emergence of expanding spacetime
by solving the classical equation of motion,
although the dimensionality of space is not determined
at the classical level.

\section{The (1+1)--dimensional expanding spacetime as an artifact of the Lorentz boost}
\label{boost}
In this section, we discuss the emergence of
the (1+1)--dimensional expanding spacetime observed
in the bosonic model, and show that this is merely an artifact of the Lorentz boost.

In the simulation, we ``gauge--fix''
the SU($N$) symmetry so that the matrix $A_0$ is diagonalized
as $A_0 = {\rm diag}(\alpha_1,\ \alpha_2,\ \cdots,\ \alpha_N)$, where
$\alpha_1 < \alpha_2 < \cdots < \alpha_N$.
In order to see the structure of the spatial matrices $A_i$ in that basis,
we define a quantity
\begin{equation}
  \mathcal{A}_{pq} = \frac{1}{9} \sum_{i=1}^9 \left| (A_i)_{pq} \right|^2 \  ,
  \label{def-calA}
\end{equation}
which is shown later in Fig.~\ref{fig:band_diagonal_after_LT_SUSY}.
As in this case, the obtained spatial matrices have a band--diagonal structure
in general when the emergent space shows a clear expanding behavior.
Using the band--width $n$, we define the time by
\begin{equation}
    t_a = \sum_{i=1}^a |\bar{\alpha}_i - \bar{\alpha}_{i-1}| \ ,
\end{equation}
where $\bar{\alpha}_i$ is an average of $\alpha$'s in the $i$--th block
with size $n$
defined as
\begin{equation}
    \bar{\alpha}_i = \frac{1}{n} \sum_{\nu=1}^{n} \Braket{\alpha_{i+\nu}} \ .
\end{equation}
We also define the $n \times n$ block matrices in the spatial matrices as
\begin{equation}
  \left( \bar{A}_i \right)_{kl}(t_a) = \left( A_i \right)_{(k+a-1)(l+a-1)} \ ,
  \label{def-barA}
\end{equation}
which are interpreted as representing
the state of the universe at $t_a$.
In what follows, we omit the index $a$
of $t_a$
for simplicity and
shift the time so that the results are
symmetric around $t=0$.
As an order parameter of the SSB of the SO(9) symmetry,
we define the ``moment of inertia tensor" as
\begin{equation}
    T_{ij}(t) = \frac{1}{n}{\rm tr}\left( \bar{A}_i(t) \bar{A}_j(t) \right) \ ,
\end{equation}
where ``tr'' is used for traces of $n\times n$ block matrices,
discriminating it from ``Tr'' used for traces of $N\times N$ matrices.
If the SO(9) symmetry is not spontaneously broken,
the nine eigenvalues $\lambda_i(t)$ of the tensor $T(t)$
become degenerate in the large--$N$ limit.
In all the simulations in this
work\footnote{In order to stabilize the complex Langevin simulations,
we have introduced a parameter $\eta$ as described in
Ref.~\cite{Anagnostopoulos:2022dak}, which is taken to be $\eta=0.005$
in this work.
This procedure is similar to the so-called dynamical stabilization, which has been
used in complex Langevin simulations of finite density
QCD \cite{Attanasio:2018rtq, Hansen:2024lkn}.},
we use $N=96,\ \gamma=4$, and the block size is chosen to be $n=12$.

\begin{figure}
    \centering
    \includegraphics[height=12em]{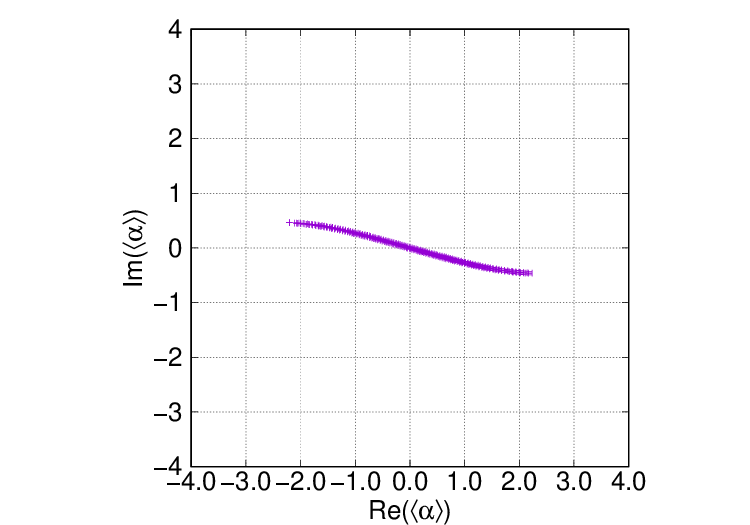}\hspace{2em}
    \includegraphics[height=12em]{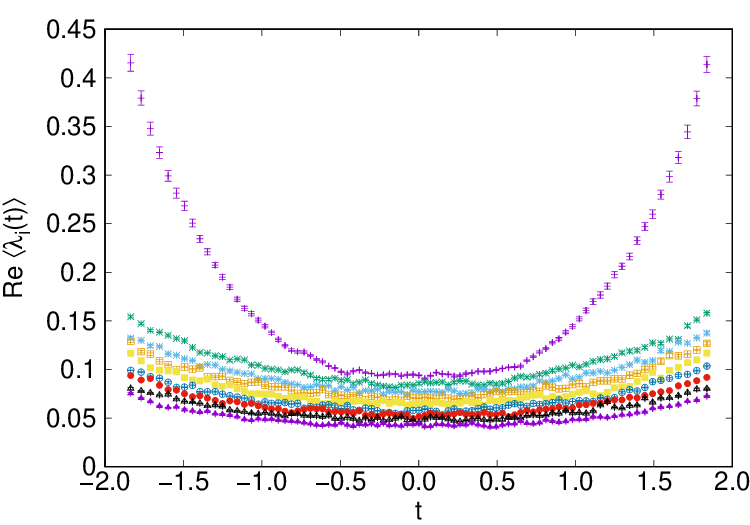}
    \caption{(Left) The eigenvalues $\alpha_i$ of the matrix $A_0$
      are plotted in the complex plane for the bosonic model with $N=96$ and $\gamma=4$.
      (Right) The real part of the eigenvalues $\lambda_i(t)$
      of $T_{ij}(t)$ are plotted against time
      for the bosonic model with $N=96$ and $\gamma=4$, where
      we find that one out of nine eigenvalues starts to grow at some point in time.}
    \label{fig:alpha_Tij_before_LT}
\end{figure}

In Fig.~\ref{fig:alpha_Tij_before_LT}, we plot the eigenvalues $\alpha_i$
of the matrix $A_0$ (Left) and the eigenvalues $\lambda_i(t)$ of $T_{ij}(t)$ (Right).
The eigenvalues $\alpha_i$ are distributed on a curve which is almost parallel to the real axis for large $|\alpha_i|$, indicating the emergence of real time at late times\footnote{We have also confirmed that the quantity 
$\langle {\rm tr} \bar{A}_i(t)^2 \rangle$
with the spatial matrices $A_i(t)$ defined by (6) is close to real,
indicating the emergence of real space at late times. 
This applies to all the cases discussed in this paper.}.
We also see that one out of nine eigenvalues of $T_{ij}(t)$ grows
with $t$,
which suggests the emergence of an expanding (1+1)--dimensional spacetime
at late times for the chosen parameters.

However, this is an artifact of Lorentz boosts as we see below.
In Fig.~\ref{fig:trace_block_before_LT} (Left),
we plot the trace of each spatial block matrix in the SO(9) basis which diagonalizes
the moment of inertia tensor $T(t)$.
We find that one of them grows linearly in time,
which indicates that the obtained configurations are Lorentz boosted.
Therefore, we need to
remove the effects of the Lorentz boost to obtain the proper information of the
emergent spacetime.

\begin{figure}
    \centering
    \includegraphics[height=12em]{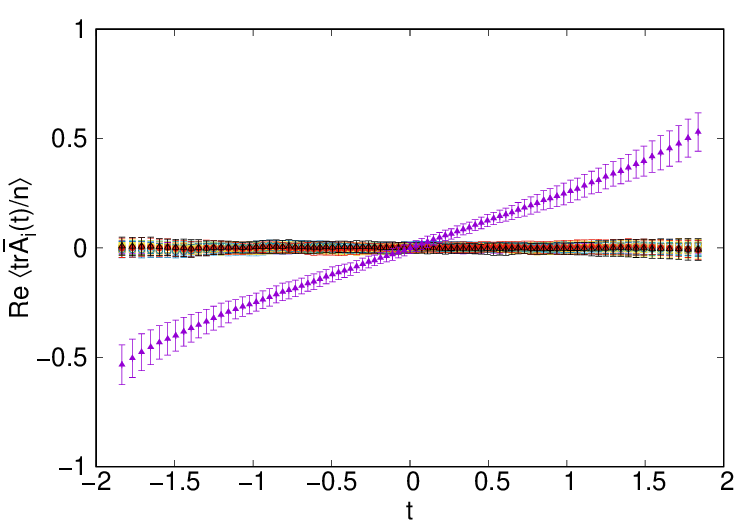}
    \hspace{2em}
    \includegraphics[height=12em]{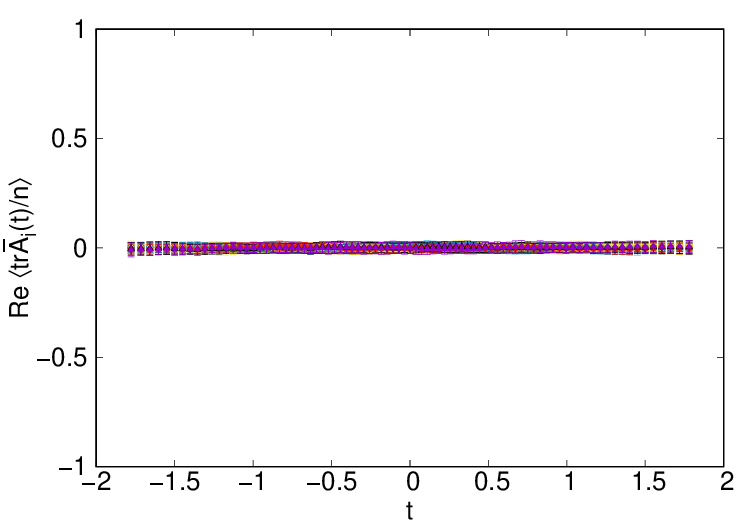}
    \caption{(Left) The real part of the trace of each spatial matrix
      is plotted against time
      for the bosonic model
      with $N=96$ and $\gamma=4$.
      Different colors of the data points correspond to a different index $i$
      for the spatial matrices $\bar{A}_i(t)$.
      (Right) The real part of the trace of each spatial matrix
      after the Lorentz transformation
      is plotted against time.
    }
    \label{fig:trace_block_before_LT}
\end{figure}

For that purpose, we choose a Lorentz frame by minimizing the quantity
\begin{equation}
    \mathcal{T} = {\rm Tr}\left(A_0^\dagger A_0 \right) \  ,
\label{min_func}
\end{equation}
with respect to Lorentz transformations on each sampled configuration.
This can be achieved by performing the (1+1)--dimensional Lorentz transformation
\begin{equation}
    \begin{pmatrix}
        A_0^\prime \\ A_i^\prime
    \end{pmatrix}
    =\begin{pmatrix}
        \cosh{\sigma} & \sinh{\sigma} \\ \sinh{\sigma} & \cosh{\sigma}
    \end{pmatrix}
    \begin{pmatrix}
        A_0 \\ A_i
    \end{pmatrix}
\end{equation}
iteratively in such a way
that the quantity (\ref{min_func}) is minimized with respect to $\sigma$ at each step,
where $i=1,2,...,9$ and $\sigma$ is a real parameter.

Let us discuss how $\sigma$ is determined at each step.
Plugging $A_0^\prime$ in Eq.~(\ref{min_func}), we get
\begin{equation}
    \mathcal{T}^\prime = \cosh^2{\sigma}~ {\rm Tr} \left( A_0^\dagger A_0 \right) + \sinh^2{\sigma}~ {\rm Tr} \left( A_i^\dagger A_i \right) + 2\cosh{\sigma}\sinh{\sigma}~ {\rm Re}~ {\rm Tr} \left( A_0^\dagger A_i \right)\  .
\end{equation}
Thus the problem reduces to the minimization of
\begin{equation}
    f(x) = a\cosh{x} + b\sinh{x} \ ,
\end{equation}
where we have defined $x=2\sigma$ and
\begin{equation}
\begin{split}
    a &= {\rm Tr} \left( A_0^\dagger A_0 \right) + {\rm Tr} \left( A_i^\dagger A_i \right) \ ,\\
    b &= 2~ {\rm Re}~ {\rm Tr} \left( A_0^\dagger A_i \right)\  .
\end{split}
\end{equation}
We find that the minimum $\sqrt{a^2-b^2}$ is
obtained at $x = \tanh^{-1}{\left(-\frac{b}{a}\right)}$,
where $\left| \frac{b}{a} \right| < 1$  as one can prove from the inequality
\begin{equation}
    {\rm Tr}\left(A_0 \pm A_i\right)^\dagger\left(A_0 \pm A_i\right) \ge 0\  .
\end{equation}

Note that the matrix $A_0$ is no longer diagonal after the Lorentz transformation.
Therefore, we redefine $\alpha_i$ by diagonalizing $A_0$ as
\begin{equation}
    A_0 \to P^{-1} A_0 P \equiv \tilde{A}_0\, ,
\end{equation}
where $P$ is a general complex matrix and $\tilde{A}_0 = {\rm diag}(\alpha_1,\ \alpha_2,\ \cdots,\ \alpha_N)$ is a complex diagonal matrix
with the ordering
\begin{equation}
    {\rm Re}~\alpha_1 < {\rm Re}~\alpha_2 < \cdots < {\rm Re}~\alpha_N\, .
\end{equation}
Accordingly, we transform the spatial matrices $A_i$ as
\begin{equation}
    A_i \to P^{-1} A_i P\, .
\end{equation}

In Fig.~\ref{fig:trace_block_before_LT} (Right), we plot the trace of each spatial
matrix after the Lorentz
transformation.  By comparing this plot with
Fig.~\ref{fig:trace_block_before_LT} (Left),
we find that the linear growth of the trace of the block matrices has disappeared.

\begin{figure}
    \centering
    \includegraphics[height=12em]{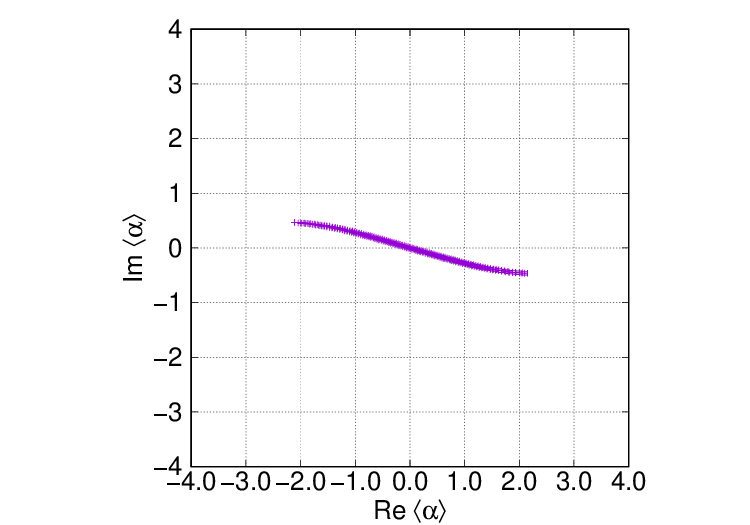}\hspace{2em}
    \includegraphics[height=12em]{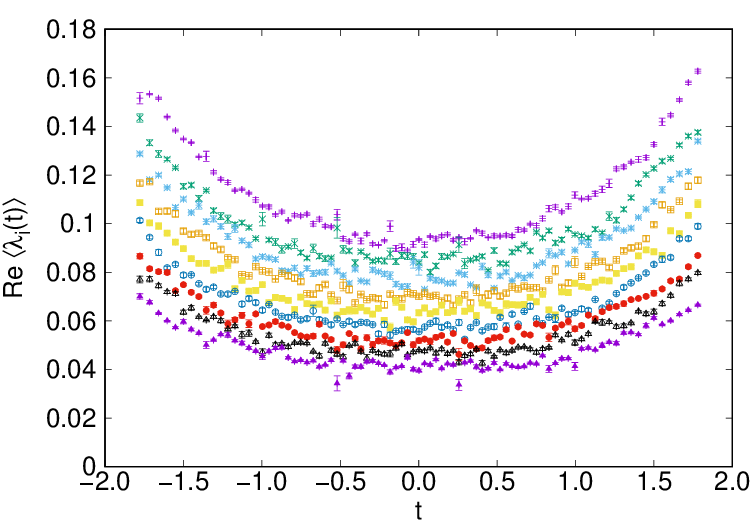}
    \caption{(Left) The eigenvalues $\alpha_i$ of the matrix $A_0$
      after the Lorentz transformation
      are plotted in the complex plane for the bosonic model with $N=96$ and $\gamma=4$.
      (Right) The real part of the eigenvalues $\lambda_i(t)$
      of $T_{ij}(t)$ after the Lorentz transformation
    are plotted against time for the bosonic model with $N=96$ and $\gamma=4$.
    The growth of one out of nine eigenvalues has disappeared.}
    \label{fig:trace_block_Tij_after_LT}
\end{figure}

In Fig.~\ref{fig:trace_block_Tij_after_LT}, we plot the eigenvalues $\alpha_i$
of the matrix $A_0$ (Left) and the eigenvalues $\lambda_i(t)$ of $T_{ij}(t)$ (Right)
after the Lorentz transformation,
which should be compared with Fig.~\ref{fig:alpha_Tij_before_LT}.
While the eigenvalue distribution of $\alpha_i$ is not affected significantly
by the Lorentz transformation,
the nine eigenvalues of $T_{ij}(t)$ come quite close to each other
after the Lorentz transformation,
indicating that the 1--dimensional expansion is indeed an artifact of the Lorentz boost.

\section{The effect of SUSY}
\label{SUSY}

\begin{figure}
    \centering
    \includegraphics[width=0.8\hsize]{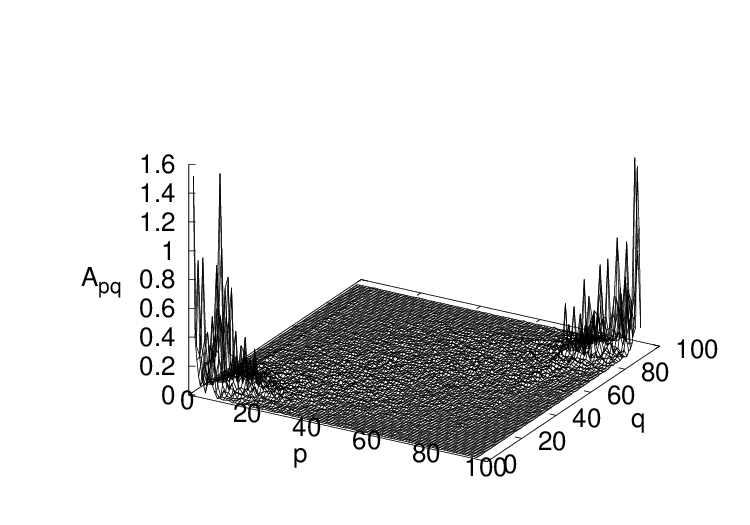}
    \caption{The structure of the spatial matrices is shown by
      $\mathcal{A}_{pq}$ defined in \eqref{def-calA}
      for the model including the fermionic contribution
      with $N=96$, $\gamma=4$, $m_{\rm f}=3.5$, $\tilde d=5$ and $\xi=16$.}
    \label{fig:band_diagonal_after_LT_SUSY}
\end{figure}

When we include the fermionic contribution, the existence of near--zero eigenvalues of the Dirac operator makes the complex Langevin
simulations untrustable.  This problem is known as
the singular drift problem \cite{Nishimura:2015pba,Nagata:2016vkn}.
In order to overcome this problem, we add a fermionic
mass term
\begin{equation}
    \label{k01}
    S_{m_{\rm f}} = iNm_{\rm f} \rm{Tr}\left[ \bar{\Psi}_\alpha \left(\Gamma_7\Gamma_8^\dagger\Gamma_9\right)_{\alpha\beta} \Psi_\beta \right],
\end{equation}
where $m_{\rm f}$ is a mass parameter.
We eventually need to make the $m_{\rm f}\to 0$ extrapolation to retrieve
the original model.
Note that, at $m_{\rm f} = \infty$, the fermionic degrees of freedom decouple and
the model becomes equivalent to the bosonic model.

We find that the complex Langevin method works only
for $m_{\rm f} \gtrsim 3.5$ with the present matrix size $N=96$.
The results for $m_{\rm f} = 3.5$, however,
are still qualitatively the same as those
of the bosonic model.
In order to enhance the
effect of SUSY
without decreasing $m_{\rm f}$,
we attempt to
suppress the bosonic fluctuations
by modifying the Lorentz invariant mass term as
\begin{equation}
    \label{k02}
    S_\gamma = \frac{1}{2}N\gamma \left\{ {\rm Tr}\left( A_0 \right)^2 - \sum_{i=1}^{\tilde{d}}{\rm Tr}\left( A_i \right)^2 - \xi \sum_{j=\tilde{d}+1}^9{\rm Tr}\left( A_j \right)^2  \right\}\ ,
\end{equation}
where $\xi\ (\ge 1)$ is an additional parameter,
which is introduced to suppress the fluctuations of $(9-\tilde{d})$ bosonic
matrices.
Note that
this term breaks the Lorentz symmetry\footnote{The idea of introducing a mass term
of this kind is inspired by a BMN--type deformation \cite{Berenstein:2002jq}
of the type IIB matrix model,
which preserves SUSY \cite{Bonelli:2002mb}. See also Ref.~\cite{Kumar:2023nya}
for complex Langevin simulations of the Euclidean model with this SUSY deformation.}
from SO(9,1) to SO($\tilde{d}$,1) for $\xi>1$.
We show our results
for this modified model with $N=96$, $\gamma=4$, $m_{\rm f}=3.5$, $\tilde{d}=5$
and $\xi=16$ after the Lorentz transformation that removes the
artifact of the Lorentz boost.

\begin{figure}
    \centering
    \includegraphics[height=12em]{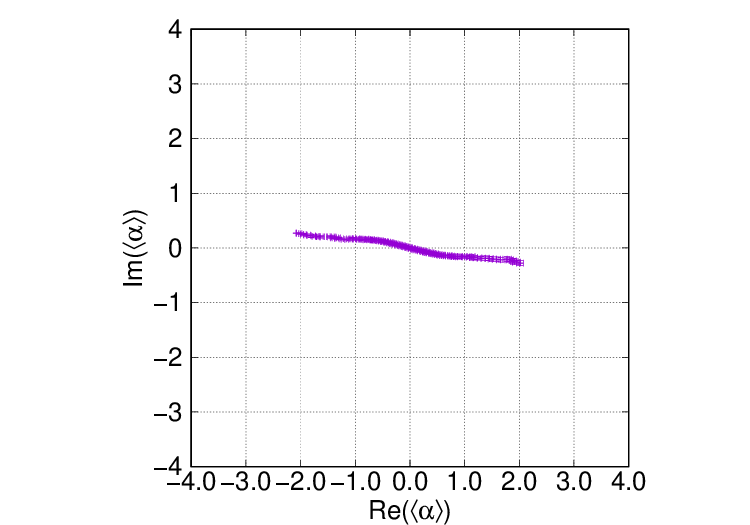}\hspace{2em}
    \includegraphics[height=12em]{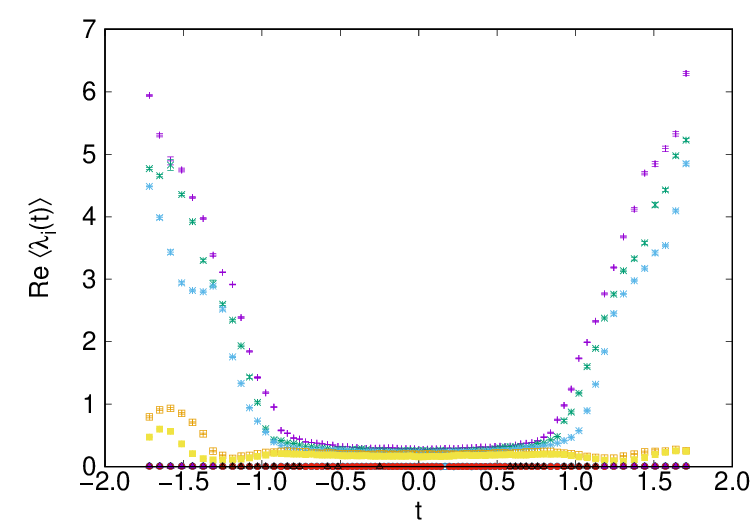}
    \caption{(Left) The eigenvalues
      $\alpha_i$ of the matrix $A_0$ are plotted in the complex plane
      for the model including the fermionic contribution
      with $N=96$, $\gamma=4$, $m_{\rm f}=3.5$, $\tilde d=5$ and $\xi=16$. 
      (Right) The real part of the eigenvalues $\lambda_i(t)$
      of $T_{ij}(t)$
      are plotted against time for the same model.      
      Three out of nine eigenvalues start to grow at some point in time, which clearly
      indicates the emergence of an expanding 3--dimensional space.}
    \label{fig:susy_alpha_eig_T}
\end{figure}

In Fig.~\ref{fig:band_diagonal_after_LT_SUSY},
we plot the quantity $\mathcal{A}_{pq}$ defined in \eqref{def-calA},
where we see
a band--diagonal structure of the spatial matrices.
In Fig.~\ref{fig:susy_alpha_eig_T} (Left), we plot the eigenvalues of $A_0$.  We find that the distribution of the eigenvalues
becomes parallel to the real axis at late times, which implies that the time becomes real in that region.  In
Fig.~\ref{fig:susy_alpha_eig_T} (Right), we plot the
eigenvalues $\lambda_i(t)$ of $T_{ij}(t)$.
While the SO($\tilde{d}$) spatial rotational symmetry seems to be
preserved at early times, it is broken spontaneously to
SO(3) at some point in time.
After this SSB, only three eigenvalues start to grow.
This result suggests that the expanding
(3+1)--dimensional spacetime appears at late times in the presence of SUSY.

\section{Summary}
\label{summary}
We have performed first--principle calculations of the
Lorentzian type IIB matrix model
using the Lorentz invariant mass term as an IR regulator.
From the simulation of the bosonic model, we found that the Lorentz boosts
may cause severe artifacts in the emergent spacetime structure.
We performed a Lorentz transformation on the sampled configurations
to remove these artifacts.
We then
found that the SSB of SO(9) does not occur in the bosonic model.
Hence, the fermionic contribution is expected to be crucial for
the emergence of (3+1)--dimensional spacetime.

When we include the fermionic contribution,
we have to modify the model by adding the fermionic mass term (\ref{k01})
in order to make
the complex Langevin method work.
We find that,
at $m_{\rm f} = 3.5$, which is the minimal value that we were able to
achieve for the present matrix size $N=96$,
the results are qualitatively the same as those of the bosonic model.

In order to enhance the effect of SUSY without decreasing $m_{\rm f}$ further,
we reduced the quantum fluctuations of the bosonic matrices
by modifying the Lorentz invariant mass term as in Eq.~(\ref{k02})
with the parameter $\xi$.
We performed simulations in the $\tilde d=5$ case and found 
that the SO($\tilde d$) spatial rotational symmetry is spontaneously broken,
and (3+1)--dimensional expanding spacetime appears at some point in
time.  Note that these results are obtained after the Lorentz transformation that
removes the artifact of the Lorentz boosts.
Recently it has been proposed \cite{Asano:2024def} that the Lorentz symmetry
should be "gauge--fixed"
in defining the Lorentzian type IIB matrix model.
Then the configurations will not get Lorentz boosted during the simulation.

In order to
investigate whether the (3+1)--dimensional spacetime emerges
in the original model, we need to take the limits of $m_{\rm f} \rightarrow 0$,
$\xi \rightarrow 1$, $N \rightarrow \infty$ and $\gamma \rightarrow 0$,  eventually.
Performing these extrapolations is an important future direction.

\section*{Acknowledgements}
T.\;A.,\ K.\;H.\ and A.\;T.\ were supported in part by Grant-in-Aid (Nos.17K05425, 19J10002, and 18K03614, 21K03532, respectively)
from Japan Society for the Promotion of Science.
This research was supported by MEXT as ``Program for Promoting Researches on the
Supercomputer Fugaku'' (Simulation for basic science: approaching the new quantum era,
JPMXP1020230411) and JICFuS.
This work used computational resources of supercomputer Fugaku provided by the RIKEN Center for Computational Science
(Project IDs: hp210165, hp220174, hp230207), and Oakbridge-CX provided by the University of Tokyo (Project IDs: hp200106, hp200130, hp210094, hp220074, hp230149), and Grand Chariot provided by Hokkaido University (Project ID: hp230149) through the HPCI System
Research Project.  Numerical computations were also carried out on Yukawa-21 at YITP in Kyoto University and on PC clusters in KEK
Computing Research Center.  This work was also supported by computational time granted by the Greek Research and Technology Network
(GRNET) in the National HPC facility ARIS, under the project IDs LIIB and LIIB2. 

\bibliographystyle{utphys}
\bibliography{bib}
\end{document}